\documentclass[prl,twocolumn,superscriptaddress,showpacs]{revtex4}

\usepackage{graphicx}

\newcommand{\etal}{{\em et al}.}

\newcommand{\EF}{$E_{\rm F}$}

\newcommand{\vcro}{(V$_{1-x}$Cr$_x$)$_2$O$_3$}

\newcommand{\vcropi}{(V$_{0.972}$Cr$_{0.028}$)$_2$O$_3$}

\newcommand{\vcropitheory}{(V$_{0.972}$Cr$_{0.038}$)$_2$O$_3$}

\begin{document}

\title{Filling of the Mott-Hubbard gap \\ in the high temperature photoemission
spectrum of \vcropi}

\author{S.-K. Mo}
\affiliation{Randall Laboratory of Physics, University of
Michigan, Ann Arbor, MI 48109}
\author{H.-D. Kim}
\affiliation{Pohang Accelerator Laboratory, Pohang 790-784, Korea}
\author{J. W. Allen}
\author{G.-H. Gweon$^{\dagger}$}
\affiliation{Randall Laboratory of Physics, University of
Michigan, Ann Arbor, MI 48109}
\author{J. D. Denlinger}
\affiliation{Advanced Light Source, Lawrence Berkeley National
Laboratory, Berkeley, CA 94720}
\author{J.-H. Park}
\affiliation{Department of Physics, Pohang University of Science
and Technology, Pohang 790-784, Korea}
\author{A. Sekiyama}
\author{A. Yamasaki}
\author{S. Suga}
\affiliation{Department of Material Physics, Graduate School of
Engineering Science, Osaka University, 1-3 Machikaneyama,
Toyonaka, Osaka 560-8531, Japan}
\author{P. Metcalf}
\affiliation{Department of Physics, Purdue University, West
Lafayette, IN 47907}
%\author{G. Keller}
%\affiliation{Theoretical Physics III, Center for Electronic
%Correlations and Magnetism, University of Augsburg, 86135
%Augsburg, Germany}
\author{K. Held}
\affiliation{Max Planck Institute for Solid State Research,
Heisenbergstrasse 1, D-70569 Stuttgart, Germany}
%\author{V. Eyert}
%\affiliation{Theoretical Physics II, University of Augsburg, 86135
%Augsburg, Germany}
%\author{V.\ I.\ Anisimov}
%\affiliation{Institute of Metal Physics, Ekaterinburg GSP-170,
%Russia}
%\author{D. Vollhardt}
%\affiliation{Theoretical Physics III, Center for Electronic
%Correlations and Magnetism, University of Augsburg, 86135
%Augsburg, Germany}

\date{Received \hspace*{30mm}}

\begin{abstract}
Photoemission spectra of the paramagnetic
insulating (PI) phase of \vcropi, taken in ultra
high vacuum up to the unusually high temperature (T) of 
$800 K$, reveal a property unique to the Mott-Hubbard 
(MH) insulator and not observed previously.  With increasing $T$ the 
MH gap is filled by spectral weight transfer, in
qualitative agreement with high-$T$ theoretical calculations 
combining dynamical mean field theory 
and band theory in the local density approximation.
\end{abstract}


\maketitle

%   Introduction

The gap of a Mott-Hubbard (MH) insulator arises from mutual electron-electron repulsions that suppress charge fluctuations \cite{Mott}, whereas the single-particle gap of a band insulator is due to coherent single electron scattering of electrons by the periodic lattice potential.  Thereby MH insulators can occur for materials where band theory would demand a metal.  
The effect of increasing temperature ($T$) is expected
to be dramatically different for insulators with the two kinds
of gaps.  Apart from thermal lifetime broadenings, the 
gap of the band insulator is not $T$-dependent, and non-zero $T$ 
merely leads to excitations of
electrons across the gap into the upper band, leaving holes in the
lower band.  By contrast, for the MH insulator the spectrum 
{\em itself} changes significantly with increasing $T$.
Specifically, the gap is filled by transferred 
incoherent spectral weight.  This counter-intuitive
and not widely appreciated property of the high-$T$ MH state 
has been reported in single-particle theoretical 
spectra \cite{Bulla} computed within  
the framework of dynamical mean-field theory (DMFT)
 \cite{DMFT, DMFTreview} applied, e.g., to the insulating state of a one-band
Hubbard model, in which an on-site Coulomb repulsion `$U$'
overcomes the bandwidth arising from site to site hopping
 `$t$' to stabilize the MH insulator.  

The primary focus of this Letter is to ascertain by 
high-$T$ photoemission spectroscopy (PES) on {\vcropi}
whether the theoretical prediction 
of a MH gap-filling holds, and to test the gap-filling
spectral shape found in a specific local density 
approximation(LDA)+DMFT calculation of the high temperature
 spectrum of PI \vcro \ \cite{Held}. A
secondary but nonetheless important aspect of the paper is to
document the difference between PI phase spectra for the bulk and
the surface regions of the sample.  We emphasize that the 
high-$T$ states appearing in the gap cannot
be described as the coherent Fermi liquid (FL) quasi-particles (QPs)
that occur at the Fermi energy \EF\ on the metallic side at low $T$.  
In the high-$T$ incoherent regime, a clear distinction between metal 
and insulator does not exist and so our study is 
directed at physics very different from that of previous studies 
of spectral changes across low-$T$ first-order metal-insulator transitions. 

\begin{figure}[!t]
\includegraphics[width=3.3 in]{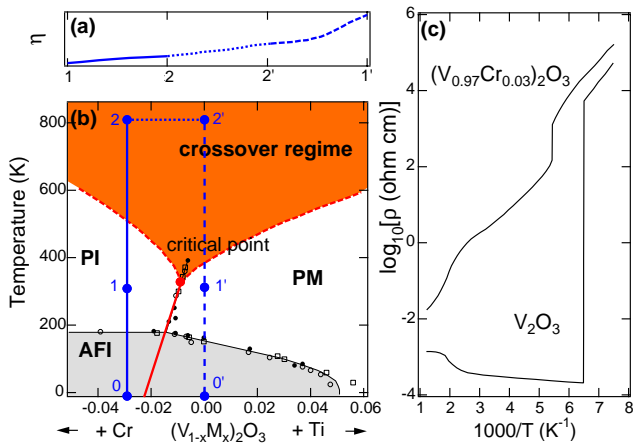}
\caption{\label{FigPD} Experimental phase diagram (panel (b)) of doped V$_2$O$_3$ \cite{McWhan} (symbols)
and DMFT (Landau) theory for the one-band Hubbard model 
(red lines). Panel (a) depicts the change of low 
energy spectral weight $\eta$ through this crossover regime along 
the path $1 \rightarrow 2  \rightarrow 2' \rightarrow 1'$. Panel (c) shows the 
temperature dependence of the resistivity  along the two
paths (taken from \cite{kuwamoto}; the small difference in doping is unimportant).
}
\end{figure}

To provide an overview for discussing both theory and experiment, 
Fig.\ \ref{FigPD} shows the experimental \vcro\ and the 
DMFT phase diagrams.  For a sketch of the latter, we 
employed \cite{footnote2} the DMFT Landau theory 
\cite{DMFTLandau} describing the behavior 
around the DMFT critical point, including theoretical 
arguments \cite{realcrit} that require the experimental 
critical point to lie {\em above} the DMFT critical point (red dot), 
as shown.  

Let us follow the changes of the low energy spectral weight, 
which is the order parameter $\eta$ of the Landau theory 
(panel (a)), along the two paths depicted in panel (b) of 
Fig.\ \ref{FigPD}.  Each path crosses the AFI phase boundary, 
producing the low-$T$ jumps seen in the resistivity (panel (c))
\cite{kuwamoto}.  Since the DMFT 
calculation does not include the AFI phase, the PM and
PI phases in the theory extend to $T=0$.
 Starting on the metallic side at $T=0$ 
(point 0'), we have a FL phase with a QP peak at \EF.
Increasing $T$ somewhat (point 1') electron-electron interactions
give rise to a non-zero FL QP lifetime $\propto T^2$ 
and the QP peak is smeared out slightly.  Along 1' to 2'
we enter the crossover regime (red) within which $\eta$ 
decreases because the metallic QP peak is dramatically smeared out 
and replaced by incoherent weight with a shallow minimum 
at \EF\ \cite{Bulla}.  The lifetime becomes too short to speak 
of a QP anymore and the high-$T$ resistivity shows a marked
deviation from its simple low-$T$ metallic behavior.

%Inserting phase diagram figure-------------------------------------
%--------------------------------------------------------------------------------

On the insulating side we have a MH gap at $T=0$ (point 0).  
Following the path of the experiment reported here, at low $T$'s (point 1)
we have a very small amount of incoherent spectral weight 
within the MH gap.  Along 1 to 2 we enter the crossover regime.  
Much more spectral weight transfers into the gap ($\eta$
is increasing) and the high-$T$ resistivity shows a marked downturn 
away from its previous simple activated behavior.
The spectrum becomes more and more like that of the PM side 
(point 2'), the distinction between the phases blurs increasingly, 
and the two resistivity curves in  Fig.\ \ref{FigPD} 
approach one another \cite{limelette}.
Thus we can think of the spectral weight transferred into the PI 
phase gap as the incoherent high-$T$ weight that replaces 
the PM phase low-$T$ QP peak.  We note that a simple gap shrinking 
could also explain the high-$T$ PI phase resistivity behavior.  
One such model \cite{goodenough} based  on a high-$T$ thermal 
variation of the crystallographic c/a ratio can 
be ruled out by the weak high-$T$ dependence of c/a that we 
have measured for our $x=0.028$ material, very close to 
that \cite{ McWhan} reported for $x=0.04$.  
In fact the PES data 
that we present next support the DMFT gap filling scenario.

%   Experiment

 We have previously reported
\cite{moprl} from PES studies of the PM phase, that the surface region
is more strongly correlated than the bulk because the reduced
coordination `$z$' on the surface reduces the band width `$zt$' on the 
surface relative to that of the bulk and renders the screening 
of $U$ to be less effective.  Studying the less correlated bulk 
requires the use of high photon energies ($h\nu$'s) with good
resolution, available only at a synchrotron, whereas experimental
constraints dictated that our high-$T$ PES be performed in a
laboratory system, for which the available $h\nu$'s render
the spectra to be more surface sensitive. Indeed an early effort 
\cite{Shin} to use low $h\nu$ PES to test the high-$T$ 
DMFT prediction of QP weight decrease in the PM phase 
could not reach a firm conclusion because the QP weight is 
too small to produce an actual peak in the low 
$h\nu$ spectrum \cite{moprl} and
because an upper $T$ limit of 300K precluded a large effect.  
Our approach is to study the correlated insulator, since the surface 
is even more strongly correlated than the bulk, and to measure 
well into the high-$T$ region suggested by Fig.\ \ref{FigPD}.

Multiple cycles of PES measurements from 300K to 800K and back to
300K were performed in a VG ESCALAB MK-II system with photons
obtained from the He I (21.2 eV) and He II (40.8 eV) lines of a He
discharge lamp.  A single-crystalline sample of \vcropi\ was
mounted on a sample holder with an embedded tungsten filament for
heating. Prior to the measurement the sample holder was degassed
at the highest temperature for a week to prevent sample
contamination by outgassing during the high temperature
measurement.  The base pressure of the vacuum system is $4 \times
10^{-11}$~Torr and during the measurement the pressure was never
higher than 2 $\times 10^{-10}$~Torr. Prior to a temperature
cycle, but not during the cycle, the sample was scraped to expose
a clean surface suitable for PES measurements.  Scraping was
chosen over cleaving to assure an angle integrated spectrum and to
allow multiple temperature cycles to be made. The spectral changes
reported here are reversible over a cycle and highly repeatable.
The Fermi level and the overall experimental resolution ($\approx
90$~meV) were determined from Fermi-edge spectra of a Mo metal
reference for each of the measured temperatures and photon
energies.  For comparison, low temperature spectra were also
obtained for cleaved surfaces in the ESCALAB system using the He
lamp and at the SPring-8 synchrotron using photons of $500$~eV.
The experimental details at SPring-8 are exactly as described
previously \cite{moprl}.

%   Figure 2

The inset of Fig. 2 shows He I and He II spectra of scraped
surfaces at $300$~K for the entire valence band region.  The V
3$d$ emission down to $\approx -3.5$~eV is well separated from the
broad O 2$p$ emission further below \EF.  As is well understood 
the part of the He I spectrum that is far from \EF\ is distorted
relative to the He II spectrum by a background that is
considerably larger and faster rising than that for He II because
of various instrumental effects involving very low energy electrons. 
The main part of Fig. 2 documents the relation between the low $T$ V
3$d$ spectrum for the photon energy of our $T$-dependent study, i.e.,
He lamp excitation, and $500$~eV photon excitation. Small differences 
in the high binding energy
tails are unimportant for the subject at hand.  The more bulk 
sensitive $500$~eV spectrum \cite{500 eV spectrum details} on a
cleaved surface shows slope changes at $\approx
-1.3$~eV and $\approx -0.7$~eV.  The same slope changes are seen
in the more surface sensitive cleaved He I spectrum, but the feature
 at $\approx -0.7$~eV has relatively less intensity. This reduced
intensity is very reminiscent of our previous finding
 \cite{moprl} that the quasi-particle \EF\ peak of the PM phase is
reduced for the more strongly correlated surface region.  Changes 
in the spectrum due to scraping, e.g. a small amount of
spectral weight observable at \EF\ due to an increased number 
of the surface states that pin the chemical potential in the
gap \cite{mu shift}, are insignificant compared to the 
remarkable temperature induced spectral changes 
that we report below.  We conclude that the conditions of our $T$-dependent 
study give spectra valid for showing the $T$-dependence of the PI phase, 
although the more strongly correlated PI surface might require
somewhat higher temperatures to be equivalent to the bulk. 

%Inserting Figure 2--------------------------------------------------------
\begin{figure}[!t]
\includegraphics[width=2.8 in]{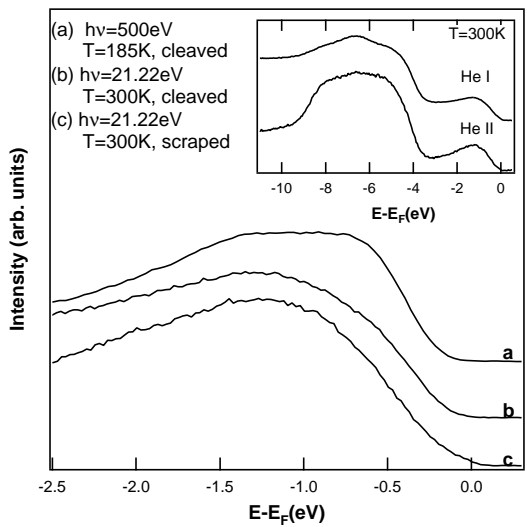}
\caption{\label{Fig 1} PES spectra of \vcropi\ measured on cleaved
surface (a) with $h\nu$=500eV, (b) with $h\nu$=21.2eV, and (c) on
a scraped surface with $h\nu$=21.2eV. The inset shows He lamp spectra
for the entire valence band.}
\end{figure}
%---------------------------------------------------------------------------------

%   Figure 3

The He II spectra in the upper panel of Fig. 3 provide the best
overview of the $T$-dependent changes.  The spectra for the two
temperatures have been normalized to match in the inelastic region
below $-10$~eV, which renders the areas under the two spectra
identical within $0.3 \%$, as should be the case. $T$-dependent
changes occur in both the V 3$d$ and the O 2$p$ regions and it
is found that the areas under each region are conserved
separately. Using this information the two He I spectra have been
normalized to preserve their areas in the V 3$d$ region, which
shows the same $T$-dependence as in the He II spectra.  The part
 of the He I spectrum farther from \EF\ shows what appears
to be (but is actually not) an extra $T$-dependence that 
interferes with observing the intrinsic $T$-dependence in 
the O 2$p$ region.  This spurious effect is typical of 
the instrumental sensitivity of this part of the He I 
background, in this case probably due to the effect 
of the magnetic field produced by the sample 
heater current in changing the paths of the very low energy electrons 
that enter the electron analyzer for this part of the He I spectrum.

%Inserting Figure 3-------------------------------------------------------------
\begin{figure}[!b]
\includegraphics[width=2.8 in]{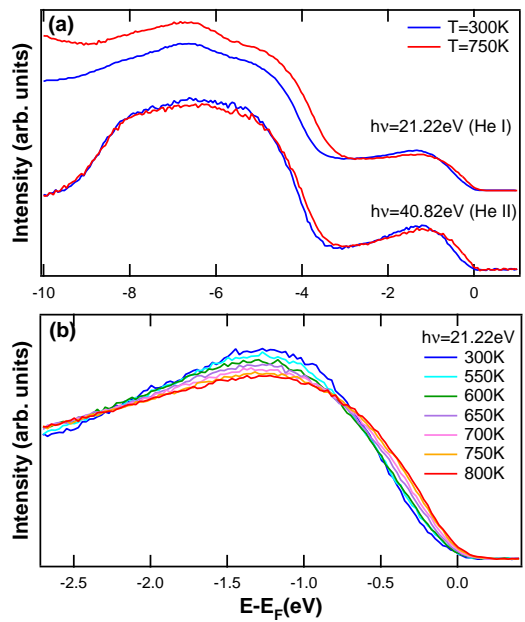}
\caption{\label{Fig 2} Temperature dependent change in the PES
spectra of \vcropi\ (a) for the entire valence band, and (b) for the
V 3$d$ region. The latter shows incremental weight transfer into the
Mott-Hubbard gap as the temperature increases.}
\end{figure}
%-------------------------------------------------------------------------------------

Our main interest in this study is the gap
region near \EF.  Since the He I spectrum shows this region well
and has much higher intensity than the He II spectrum, detailed He
I spectra were taken in the gap region at 50 K intervals.  These
spectra, presented in the lower panel of Fig. 3, show a
substantial change, smoothly progressing, over the temperature
range studied.  Beginning at $550$~K, roughly consistent 
with entering the crossover regime along path $1 \rightarrow 2$ in
Fig. 1, spectral weight is transferred steadily into the gap 
from a region as much as $1.5$~eV below the gap,
causing the curvature of the spectral onset to change
from positive to negative over the first 0.7 eV.  This spectacular
change is entirely different from what would be expected for thermal
generation of holes below a band insulator gap and cannot be 
understood as an ordinary thermal broadening or a simple phonon effect, 
as we discuss next.

%Inserting Figure 4----------------------------------------------------------
\begin{figure}[!t]
\includegraphics[width=2.6 in]{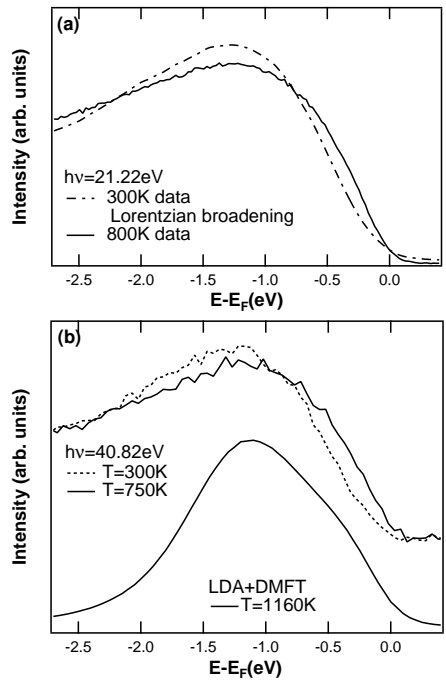}
\caption{\label{Fig 3} (a) Comparison to show that Lorentzian 
broadening of the 300 K spectrum, even by as much as 1160K, 
cannot account for the thermal change in the 800K spectrum.  
(b) He II PES spectra measured at 300K and 750K compared with the high temperature LDA+DMFT calculation \cite{Held}, including the experimental resolution.}
\end{figure}
%-----------------------------------------------------------------------------------

% Figure 4

We now summarize the case that the observed
thermal change is best understood as the gap filling expected 
in theories of the MH insulator.  We note immediately that the 
transfer energy range of $1.5$~eV vastly exceeds phonon 
energies and is 20$\times$k$_{B}T$ for $T=800$K.  The upper 
panel of Fig. 4 compares the 800K spectrum to an overestimate 
of simple thermal broadening of the 300K spectrum, obtained by 
broadening with an 1160K Lorentzian.  Such thermal
broadening does not account for the spectral weight transfer 
and the reversed curvature of the 800K spectrum. The spectral 
evolution is also quite different from that of a simple gap 
closing, for which changes primarily near the chemical 
potential would be expected.  The lower 
panel of Fig. 4 again shows the experimental 
change, using this time the He II spectra at low and
high $T$, along with a theoretical LDA+DMFT V 3$d$ spectrum 
for the PI phase of \vcropitheory\ at 1160K, taken 
from the calculation of Ref. \cite{Held} for a $U$ value of $5.0$~eV. 
The Cr doping is one for which the detailed crystallographic
information needed for the calculation is available and the
difference from that of our experiment is not
important for the qualitative comparison made here.
The theory reproduces the general gap filling shape, including the 
unusual negative curvature seen in the data over the first 0.7 eV.  
We conclude that the cross-over regime of the DMFT provides a
unified picture of both the high-$T$ resistivity behavior and
our high-$T$ PI phase PES spectra.

%   Summary

In summary, unusually high-$T$ PES measurements show for the PI 
phase of \vcro\ a previously unobserved transfer of spectral 
weight into the gap region, well correlated with the high-$T$ 
resistivity behavior.  We interpret our data as the first
observation of the incoherent low energy weight that defines
the cross-over regime of the DMFT MH phase diagram.

%\acknowledgements
\begin{acknowledgments}
This work was supported by the U.S. NSF at the University of
Michigan (UM) (Grant No.~DMR-03-02825), by a Grant-in-Aid for
COE Research (10CE2004) of
MEXT, Japan, by JASRI (No.~2000B0335-NS-np), by KOSEF through eSSC
at POSTECH and by the Emmy-Noether program of the Deutsche 
Forschungsgemeinschaft.  We are grateful to G. Keller and D. Vollhardt
for valuable discussions.
\end{acknowledgments}

$^{\dagger}$ Current address: Lawrence Berkeley National Laboratory,
Advanced Light Source, MS 7-100, 1 Cyclotron Road, Berkeley, CA, 94720.


\begin{thebibliography}{99}
\bibitem{footnote2} Both PM and PI solutions, one metastable, 
coexist below the DMFT critical point 
(set at $T_c\!=\!317$\rm{K}, $x_c\!=\!0.0092$ in Fig.\ 1).  The Landau 
theory describes the behavior around $T_c$ 
in terms of an order parameter $\eta$ which is given by 
$[p_x(x-x_c)+p_T(T-T_c)]\eta+\eta^3\!=\!h_x(x-x_c)+h_T(T-T_c)$ 
\cite{DMFTLandau}.  To reproduce the slope of the experimental 
first-order line and the general DMFT behavior \cite{DMFTreview}, 
we used $h_x\!=\!-2.43$, $h_T\!=\!-0.1/1000K$, $p_x\!=\!0.61$, and $p_T\!=\!0.6/1000K$.
The boundary of the crossover regime 
is not sharply defined.  For the line shown in Fig.\ 1, we used 
the criterion that $d \eta/dx$ is becoming small, i.e.,
$25\%$ of its maximal value at the same $T$.
\bibitem{500 eV spectrum details} A small contribution to the
$500$~eV spectrum from Cr 3$d$ emission has been removed by using
information obtained from resonant photoemission spectra at the V 2$p$
edge, for which the Cr emission is entirely suppressed, and from
PI phase spectra for a much smaller value of $x$.
\end{thebibliography}
\end{document}